\documentstyle[preprint,aps,pre,epsfig]{revtex}

\draft

\begin{document}
 
\title{Structures and propagation in globally coupled systems
\\ with time delays}

\author{Dami\'an H. Zanette}

\address{Consejo Nacional de
Investigaciones Cient\'{\i}ficas y T\'ecnicas\\ Centro At\'omico
Bariloche and Instituto Balseiro, 8400 Bariloche, Argentina}

\date{\today}

\maketitle

\begin{abstract}
We  consider an ensemble  of globally coupled  phase oscillators whose
interaction  is transmitted at  finite   speed.  This introduces  time
delays, which make  the spatial coordinates relevant  in  spite of the
infinite range of the interaction. We show that one-dimensional arrays
synchronize in an asymptotic state where all  the oscillators have the
same   frequency, whereas  their  phases are   distributed in  spatial
structures that --in the  case of periodic boundaries-- can propagate,
much as in coupled systems with local interactions.
\end{abstract}


\newpage

\section{Introduction}

Standard models for  studying collective complex behavior in  natural
systems consist  typically  of   ensembles of  interacting   dynamical
elements  \cite{Mikh}. Such kind of models  has proven to be extremely
versatile in  the   mathematical  description, both    analytical  and
numerical,  of  a wide   variety  of phenomena   within the scopes  of
physics, biology, and  other branches of  science \cite{KK}. According
to the range of the involved interactions, these models can be divided
into two  well   distinct  classes.  Local  interactions  --which  are
paradigmatically  represented    in       reaction-diffusion   systems
\cite{Mikh}--  give rise to   macroscopic  evolution  in which   space
variables play   a  relevant  role, such   as  spatial structures  and
propagation phenomena.  On  the other  hand, with global  interactions
--where the  coupling range is of  the order of,  or larger  than, the
system   size-- space becomes irrelevant   and collective behavior is
observed to develop in time, typically, in the form of synchronization
\cite{Kur}.

An essential model of globally coupled  elements is given  by a set of
$N$    identical  oscillators  described,   in    the  so-called phase
approximation, by  phase variables $\phi_i(t)$ ($i=1,\dots,N$).  Their
evolution is governed by the equations  
\begin{equation}
\dot \phi_i = \omega+\frac{\epsilon}{N} 
\sum_{j=1}^N \sin (\phi_j-\phi_i). 
\end{equation}
It  is known that, for any value
of the coupling  intensity $\epsilon$, all the  elements converge to a
single orbit   whose  frequency $\omega$ coincides  with  that   of an
individual  oscillator  \cite{Kur}.   In this  case,  $\epsilon^{-1}$
measures the time required to reach such synchronized state.

In this note we present results  on  a generalization of  the
above  model, where time  delays are  introduced.   The effect of time
delays in synchronization   phenomena has already  been considered for
two-oscillator   systems,   both periodic   \cite{peri}   and  chaotic
\cite{chao}. Ensembles with local interactions \cite{loc} and globally
interacting inhomogeneous systems have  also been studied \cite{2000}.
None of these  contributions make however   explicit reference to  the
relevant case where    interactions are global but   their propagation
occurs  at a  finite  velocity $v$.   This situation, which  naturally
introduces   time delays, provides  a  realistic description of highly
connected systems where   the time scales   associated with individual
evolution    and    with       mutual   signal    transmission     are
comparable. Instances  of  such   systems are neural    and informatic
networks \cite{net}, and biological  populations with  relatively slow
communication  media --such as sound  \cite{biol}. Our  main result is
that, since a finite signal  velocity makes spatial variables relevant
even  when interactions are   global, globally coupled  ensembles with
time    delays exhibit typical  features  of  systems  driven by local
interactions, in particular, structure formation and propagation.

\section{The model and its solution for short delays}

We consider an ensemble of $N$ identical oscillators in the phase
approximation, governed by the equations
\begin{equation} \label{ens}
\dot \phi_i(t) = \omega+{\epsilon \over N}
\sum_{j=1}^N \sin [\phi_j (t-\tau_{ij})-\phi_i(t)],
\end{equation}
where $\tau_{ij}=d_{ij}/v$ is  the  time required  for the signal   to
travel from element $j$  to element $i$  at velocity $v$, and $d_{ij}$
is the   distance between $i$ and  $j$.  Note that  coupling  is still
global, since its intensity $\epsilon$ does not depend on the distance
between  elements.  However, the relative  position of the oscillators
becomes now relevant through time delays.

The full specification of our system requires to  fix the topology and
the metric properties  of the ensemble, by  fixing the values $d_{ij}$
for all $i,j=1, \dots ,N$.   Moreover, initial conditions for $\phi_i$
must be provided. In the case  of delay equations like (\ref{ens}), it
is necessary to specify  the evolution of  $\phi_i$ at times prior  to
$t=0$ up to  a time $T_i=-\max \{\tau_{ij}  \}_j$ \cite{delay}. In the
following  we  shall  assume that for  $t<0$  the   oscillators evolve
independently from  each other at  their proper frequency $\omega$ and
with  random  relative    phases.    Namely,   for  $t<0$ we      have
$\phi_i(t)=\omega t+\phi_i(0)$, where  $\phi_i(0)$ is chosen at random
from  a uniform distribution   in  $[-\pi,\pi)$. At $t=0$ coupling  in
switched  on,  so  that  we formally   have  a time-dependent coupling
intensity  $\epsilon(t)= \epsilon  \theta(t)$,  where  $\theta$ is the
Heaviside step function.

Through extensive numerical calculations for  a variety of topologies,
ranging from  one-dimensional arrays to tree (ultrametric) structures,
we  have found  that  the system  evolves  to  a state  where  all the
oscillators  have the same frequency. On  the other  hand, in contrast
with  the case  without time delays  \cite{Kur},  their phases  can be
different. This asymptotic state corresponds  thus  to a situation  of
frequency synchronization.  The long-time evolution of each oscillator
can then be written as  $\phi_i(t)=\Omega t+\psi_i$, where in  general
$\psi_i\neq \psi_j$   for $i\neq j$.  The  fact that  these phases are
different could  have been expected for topologies  where not  all the
elements  are    equivalent --for    instance,  when   boundaries  are
present. As  we  show later, however,  such  states are  also found in
homogeneous topologies. In  this    case, they are   associated   with
propagating structures.

In general, the synchronization frequency is different from the proper
frequency  of   each oscillator,  $\Omega \neq  \omega$.  According to
(\ref{ens}), the synchronization frequency satisfies
\begin{equation} \label{Omega}
\Omega = \omega-{\epsilon\over N}\sum_{j=1}^N \sin(\Omega \tau_{ij}-
\psi_j+\psi_i).
\end{equation}
Note  that  the  sums $S_i=\sum_j \sin(\Omega\tau_{ij}-\psi_j+\psi_i)$
are in general different for each  $i$. However, their numerical value
must  coincide  if  the   synchronization  frequency  is   to be  well
defined. For a given value of $\Omega$, this constraint provides $N-1$
independent equations for the phases $\psi_i$:
\begin{equation} \label{s=s}
S_1=S_2=\dots =S_N.
\end{equation}
Since phases are defined up to an additive constant  we can choose for
instance $\psi_1=0$,   and  solve these  equations  for  $\psi_2,\dots
\psi_N$.   Then,   $\Omega$   can be    found self-consistently   from
(\ref{Omega}).  For large  values  of $N$,  and  due  to the  involved
nonlinearities, this results to be a quite hard numerical problem.

An approximate  solution can however been found  in the case  of short
delays,   i.e. close   to  the  situation  where  the  system is  also
synchronized in phase,  $\psi_i=\psi_j$  for all $i,j$.  Assuming that
$|\Omega \tau_{ij}-\psi_j+\psi_i| \ll 1$, we can write
\begin{equation}
S_i\approx \sum_j (\Omega\tau_{ij}-\psi_j+\psi_i)=
N\Omega\langle \tau_i \rangle-\sum_j\psi_j +N\psi_i,
\end{equation}
where $\langle \tau_i\rangle=N^{-1}\sum_j \tau_{ij}$ is the average of
the time delays associated  with   element $i$. Taking  into   account
Eq. (\ref{s=s}), we get
\begin{equation} \label{appr}
\psi_i\approx \Psi-\Omega\langle \tau_i\rangle.
\end{equation}
where  $\Psi$ is a  constant, independent of $i$,   that can be chosen
arbitrarily. This result indicates that,   in this short-delay  limit,
oscillators  with small average  delays   are relatively ahead in  the
evolution, as their phases  are larger. This  is plausibly due  to the
fact that,  in average,  they receive  the  information on  the system
state before other   elements  with larger values  of  $\langle \tau_i
\rangle$, which are thus relatively retarded.  Note moreover that in a
homogeneous topology all the elements are equivalent, so that $\langle
\tau_i\rangle$ is the  same for  all  oscillators. In  this case,  the
system is also synchronized in phase.

Within the approximation of short delays, the synchronization 
frequency is given by
\begin{equation}
\Omega \approx \frac{\omega}{1+\epsilon\langle\langle\tau
\rangle\rangle},
\end{equation}
where $\langle \langle \tau\rangle\rangle =N^{-1} \sum_i\langle \tau_i
\rangle$ is the overall    average delay. It therefore  results   that
$\Omega$ is smaller than the proper frequency of each oscillator.

\section{One-dimensional arrays}

In this note, we specifically focus  the attention on the case
of one-dimensional  arrays.  Two different  topologies are considered,
namely, with periodic boundary conditions --where all the elements are
equivalent-- and with free boundaries --where the neighborhood of each
element depends on   its distance to   the  center of the array.   For
periodic boundary conditions, which   we consider first,  the distance
between   two  elements is  not  univoquely  defined, since  it can be
measured around the  ring  in both  directions.   We fix   $d_{ij}$ by
taking the minimum    of  these  values,  namely, $d_{ij}=   \min   \{
|i-j|,N-|i-j|\}$. The delay time is  thus  $\tau_{ij}= \tau_0 \min  \{
|i-j|,N-|i-j|\}$,  where $\tau_0$ is the  time required for the signal
to travel between nearest neighbors.

In equations (\ref{ens}), the proper frequency $\omega$ can be used to
define time units   so that,  without  loss   of  generality, we   fix
$\omega=1$. Moreover, our  numerical simulations are restricted to the
case  $\epsilon=1$. As  a matter of  fact,  we  have found that  other
coupling   intensities   do   not   produce qualitatively    different
results. Note that this would  not be the  case if the oscillators had
chaotic individual dynamics.      In  such situation,   the  value  of
$\epsilon$  is   expected to control   the   existence of synchronized
states.

We have solved  numerically   equations (\ref{ens}) for  ensembles  of
$N=10^2$  to  $10^4$  oscillators with    a standard finite-difference
scheme. For small values of $\tau_0$ we find  that the above described
random-phase initial conditions    evolve to a state   of synchronized
frequency     where   the   phases   of     all  oscillators coincide,
$\psi_i=\psi_j$  for all $i$,  $j$.  This fully synchronized state  is
completely analogous to that of globally coupled identical oscillators
without time  delays,   and corresponds  to  the approximate  solution
(\ref{appr}) for the present  homogeneous   topology.  In this   case,
(\ref{Omega}) becomes an autonomous equation for $\Omega$.  The sum in
the right-hand side can in  fact be explicitly evaluated --though  its
expression depends on $N$ being even  or odd-- and the synchronization
frequency can be found  numerically by standard methods.   In general,
this equation admits  several solutions.  For  the  values of $\tau_0$
where  the  state of phase  synchronization   is encountered, however,
there is only one possible value of $\Omega$.

At $\tau_0  \approx 5 N^{-1}$ a  qualitative change occurs. Above this
critical value, the asymptotic synchronized state is not characterized
by  a  homogeneous phase anymore.  Instead,  the phase varies linearly
along the system, in such a way that a  phase difference $\Delta \psi=
\pm 2\pi$ accumulates  in a  whole  turn around.  The sign of  $\Delta
\psi$ is defined by the initial condition. Symmetry considerations, in
fact, indicate that both  signs will be  found with  equal probability
over the   set  of  initial conditions    that lead  to  this  kind of
asymptotic  state. The individual  phases are given by $\psi_i=\psi_0+
i\delta \psi$, with $\delta\psi= \pm 2\pi/N$ and $\psi_0$ an arbitrary
constant.  Due to  the time  evolution  of $\phi_i(t)=\Omega t+i\delta
\psi +\psi_0$ a structure propagates around the system at velocity
$V_1=-\Omega/\delta \psi$.
 
\begin{figure}
\begin{center}
\psfig{figure=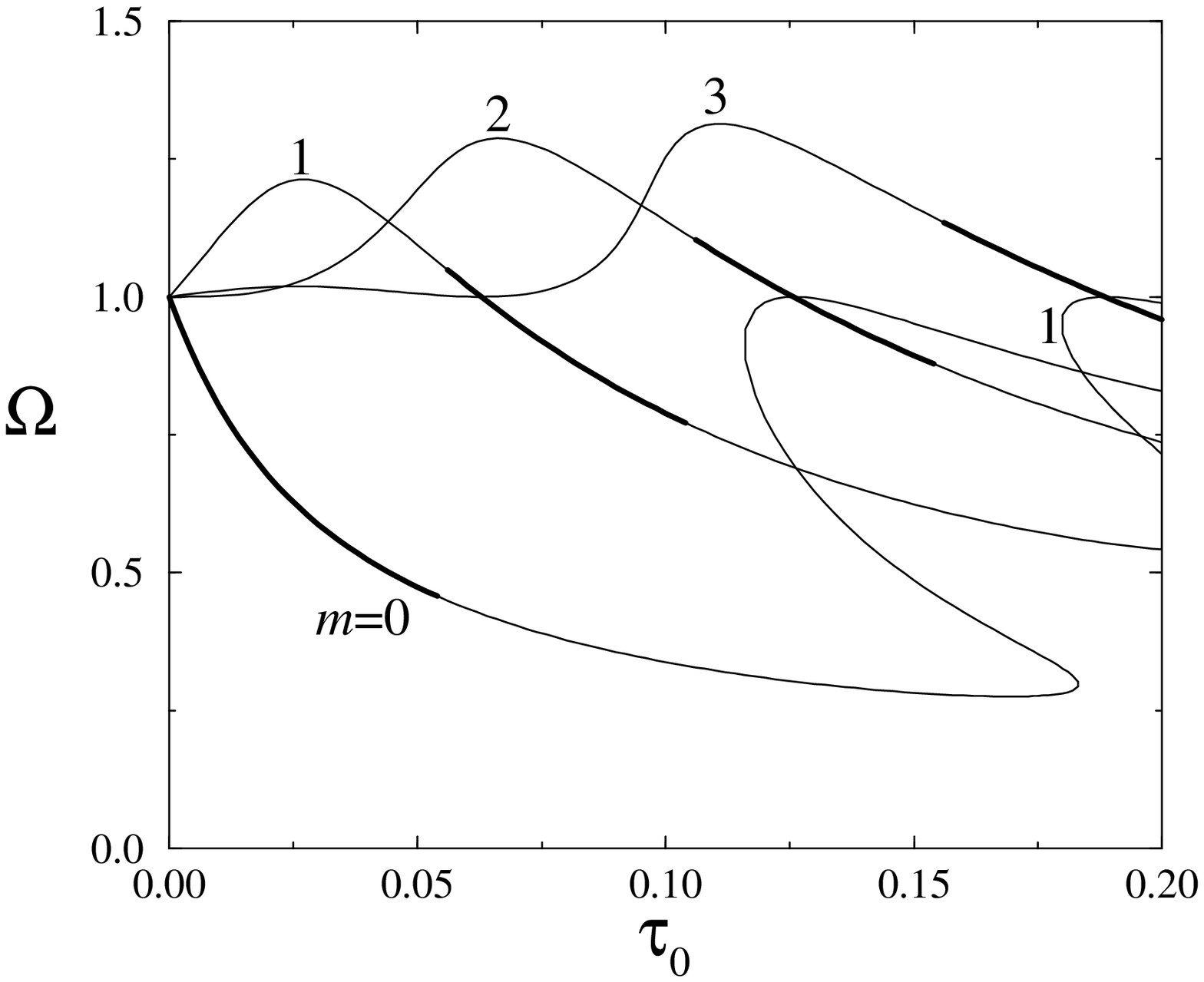,width=10 cm}
\end{center}
\caption{Synchronization frequency $\Omega$   of the asymptotic  modes
$m=0,\dots,3$ in  a   one-dimensional  ensemble  of $N=100$   globally
coupled oscillators with  periodic boundary conditions, as a  function
of the nearest-neighbor delay time $\tau_0$.}
\label{f1}
\end{figure}

Similar qualitative changes are   found at larger values of  $\tau_0$.
For $\tau_0 \approx  11N^{-1},16N^{-1}, \dots$, the asymptotic  states
modify their phase structure  in such a way  that the phase difference
around the whole system, $m\Delta \psi=\pm 2\pi m$ with $m=2,3,\dots$,
increases progressively.  The  corresponding individual   evolution is
$\phi_i(t)=  \Omega    t+im\delta    \psi+\psi_0$,  which  defines   a
propagation  velocity  $V_m=-\Omega/m\delta \psi$. The synchronization
frequency is given by
\begin{equation} \label{Omegam}
\Omega = \omega-{\epsilon\over N}\sum_j \sin [\Omega \tau_0
\min \{ |i-j|,N-|i-j| \} +(i-j) m \delta \psi].
\end{equation}
For $m=0$  this reduces to the case  of full synchronization found for
small  $\tau_0$.   Figure \ref{f1}  shows the  solutions   of equation
(\ref{Omegam}) for  various values of $m$,  and $N=100$. Bolder curves
indicate the  intervals   where each  mode has   been observed in  the
numerical calculations with  random-phase initial conditions. Note the
zones where more than one solution exist for $m=0$ and $m=1$.

Are  the transitions observed at the  above  quoted values of $\tau_0$
actual  bifurcations, associated with changes in  the stability of the
asymptotic states?  In view  of the  difficulty   of dealing  with the
linear   stability problem  for  a  many-dimensional  system with time
delays  such as (\ref{ens})  \cite{delay},  we  choose to answer  this
question  by  numerical  means.   For a  given   value  of $\tau_0$ we
calculate the  frequency $\Omega$ of   a given mode $m$  from equation
(\ref{Omegam}) and generate an  initial condition which corresponds to
that mode   added  with    a  certain   --typically    random--  small
perturbation.  Then, we run  the evolution  and  study the  asymptotic
behavior.  This has   been  carried out for $m=0,\dots,3$   at several
values  of $\tau_0$ in  ($0,0.2$), for a  100-oscillator ensemble.  In
almost  all cases,   it  has been found   that for  sufficiently small
perturbations   the considered  states are   stable for   any value of
$\tau_0$. The only exceptions seem to  be the states whose frequencies
are multiple solutions of equation (\ref{Omegam}),  since in this case
the only stable state correspond to the smallest frequency.

The  observed   transitions are therefore   not   related to stability
changes in the propagation   modes. Rather, several modes coexist  and
the   system  is multistable.  The specific  asymptotic  state is thus
selected by the initial condition. The fact that from the random-phase
initial conditions considered previously  the system evolves to a well
defined synchronous  mode,   whose  order $m$   grows   with $\tau_0$,
suggests  that the attraction basins  of  the various solutions  could
considerably   vary  in  size as  $\tau_0$    changes. Indeed, from  a
probabilistic  viewpoint,    most  initial conditions  are    of   the
random-phase type. Initial  conditions  that,  for a given   value  of
$\tau_0$,  do  not evolve to  the   mode marked  with  a bold  line in
Fig. \ref{f1} should be considered probabilistically rare.

We  consider  now  the case  of    a one-dimensional  array  with free
boundaries. Here, the distance between elements can  be defined in the
standard form, $d_{ij}=|i-j|$, so that  the time delays are $\tau_{ij}
= \tau_0 |i-j|$. In this topology sites are not equivalent. Delays for
elements near  the center of the  array are in  average lower than for
elements towards the   boundaries.  As a consequence, no   homogeneous
stable states are expected for the coupled ensemble. Numerical results
show that,  in fact, in the asymptotic  evolution  all the oscillators
have the same frequency, given by
\begin{equation} \label{Omega2}
\Omega = \omega-{\epsilon\over N}\sum_j \sin (\Omega \tau_0
|i-j| -\psi_j+\psi_i),
\end{equation}
but $\psi_i \neq  \psi_j$ if $i\neq j$  for any nearest-neighbor time
delay  $\tau_0$.   Unexpectedly,  however,  the    associated  spatial
structures not always preserve the topological symmetry of the system,
as shown in the following.
 
\begin{figure}
\begin{center}
\psfig{figure=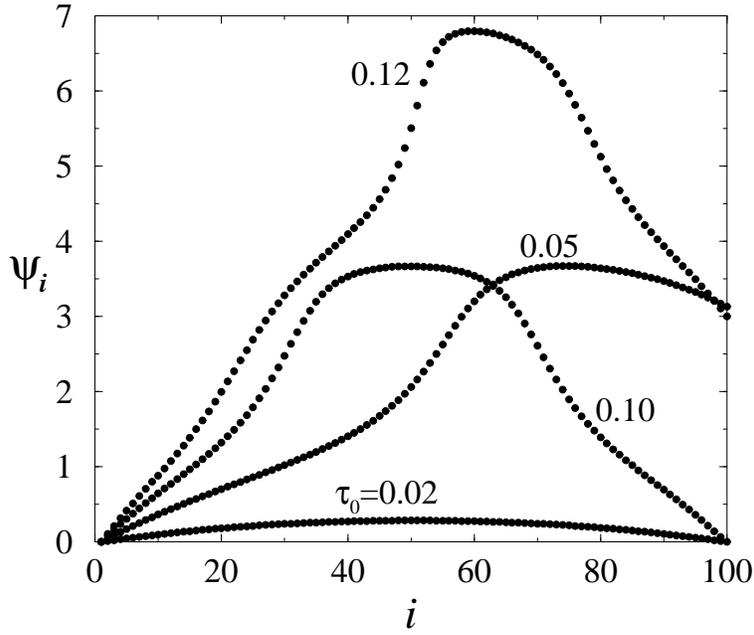,width=10 cm}
\end{center}
\caption{Stationary phase patterns in a one-dimensional 100-oscillator
array  with free boundaries, for  various  values of $\tau_0$. Without
loosing generality, we have fixed $\psi_1=0$.}
\label{f2}
\end{figure}

Our numerical calculations for  the case of free boundaries correspond
to a 100-oscillator ensemble with  the random-phase initial conditions
described  above.  For  small values of  $\tau_0$  we find a symmetric
asymptotic pattern,   $\psi_i   = \psi_{N/2-i}$, where   the   central
elements have  larger phases than near  the  boundaries (Fig. \ref{f2}
for  $\tau=0.02$). This    structure corresponds  to  the  approximate
solution (\ref{appr}) which,  for this topology,  predicts a parabolic
phase profile with  a maximum at  the center of  the array.  Beyond  a
critical value $\tau_0 \approx  0.025$ random-phase initial conditions
are instead attracted towards an asymmetric structure, where the phase
varies in $|\psi_N-\psi_1|\approx \pi$ from one end  to the other, and
attains a maximum in between. Figure \ref{f2} shows such structure for
$\tau_0   = 0.05$.  In average, of   course,  half of the realizations
produce the  symmetric   counterpart of  this  asymptotic  state.  The
situation  changes again at $\tau_0\approx  0.06$.  Beyond this point,
stationary structures  are again symmetric, as  shown in Fig. \ref{f2}
for $\tau_0=0.1$.  They result however to be more complicated than for
small $\tau_0$, with  inflection points at $i\approx  N/4$ and  a much
flatter maximum. A new critical point occurs at $\tau_0 \approx 0.11$,
where phase structures become asymmetric once more (see Fig. \ref{f2},
for $\tau_0 =0.12$).  The phase variation  between the ends is similar
to that observed for smaller $\tau_0$ but the intermediate geometry is
considerably more complex.

An analytical or semi-analytical study of these structures --including
their  existence  and stability properties--  requires considering the
consistency    problem   discussed      in   connection    with    Eq.
(\ref{Omega}). Fixing  $\psi_1=0$,  the  $N-1$ equations for  $\psi_i$
($i=2,\dots,N$) read here
\begin{equation} \label{S}
\sum_j \sin (\Omega \tau_0 |i-j|-\psi_j+\psi_i)=
\sum_j \sin[ \Omega \tau_0 (j-1)-\psi_j].
\end{equation}
This problem   will be discussed in  detail  in a separate publication
\cite{forth}. Let us stress for the moment that, though the appearance
of spatial structures was to be expected in an inhomogeneous system as
the present one-dimensional array with free boundaries, these patterns
are found to exhibit an unexpected richness upon variation of $\tau_0$
--including, in particular, symmetry breaking.

\section{Summary and discussion}

We  have  found   that  an  ensemble   of   identical globally coupled
oscillators  with finite  interaction velocity, which  gives origin to
time delays, evolves to an asymptotic  state where all the oscillators
have   the  same  frequency  but  different   phases.   Generally, the
synchronization frequency   differs  from  the  proper   frequency  of
individual  oscillators, so that the dynamics  of  each element in the
collective asymptotic  motion  does not coincide  with  its individual
(uncoupled) dynamics. Phases,  in  turn, are distributed  according to
spatial   patterns    with  nontrivial     topological   and dynamical
properties. Specifically, in a  one-dimensional periodic array several
asymptotic states coexist,  corresponding  to propagation  modes  with
different  velocities. In  a   bounded one-dimensional  array we  have
observed stationary  phase structures whose symmetry properties depend
on the size  of time delays.  These features, which are reminiscent of
the  behavior of  reaction-diffusion  systems with local interactions,
point out  sharp differences  with  the collective motion  of  coupled
oscillators without time delays.

\begin{figure}
\begin{center}
\psfig{figure=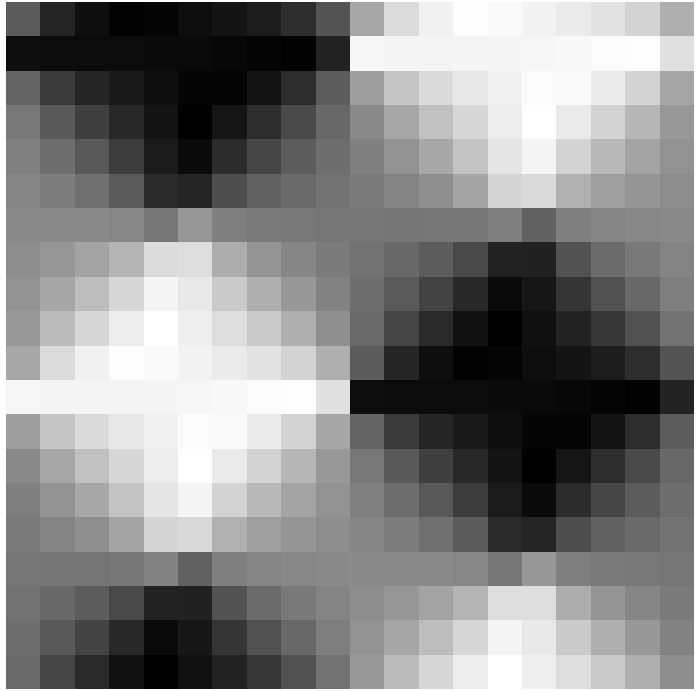,width=10 cm}
\end{center}
\caption{Snapshot   of a propagating  structure   in a two-dimensional
$20\times  20$-oscillator  array with  periodic   boundary conditions.
Dark and light  zones correspond to   phases near zero and  $\pm \pi$,
respectively.}
\label{f3}
\end{figure}

It is natural to ask whether any  structure similar to those described
above is  observed in other, more  complex  topologies.  To advance an
answer to this question, we have performed a preliminary analysis of a
two-dimensional array of $20\times 20$ elements with periodic boundary
conditions.  In this case, each element can be labeled by two indices,
$i_x$  and $i_y$,  according  to  its  Cartesian   coordinates in  the
lattice.  For algorithmic  convenience  we have defined  the  distance
between  elements as $d_{ij}       = ||i-j||_1= \min\{      |i_x-j_x|,
L-|i_x-j_x|\} +\min\{  |i_y-j_y|, L-|i_y-j_y|\}$, with  $L=20$ in  our
case.    As above, the delay   time  is $\tau_{ij}=\tau_0 d_{ij}$.  In
complete agreement with the one-dimensional analog, we have here found
that for small $\tau_0$ the system  synchronizes both in frequency and
phase.  Beyond a critical    value $\tau_0 \approx   0.025$,  instead,
propagating phase patterns  are  observed. Figure  \ref{f3} shows  the
simplest of these patterns, corresponding to the propagation mode with
$m_x=m_y=1$.

Work in progress is being devoted to  the detailed characterization of
the phase structures described  in this note, as well as those
that could arise  in other topologies. The  next step will be to study
the effects of the present kind of  time delays in ensembles formed by
chaotic   oscillators,  where  coupling   competes   as a  stabilizing
mechanism against  the   inherently  unstable  dynamics of  individual
elements.

\section*{Acknowledgment}

This work was partially carried out at the Abdus Salam International
Centre for Theoretical Physics. The author wishes to thank the Centre
for hospitality.

\newpage

\end{document}